\documentclass[aps,twocolumn,superscriptaddress]{revtex4-1}
\usepackage[T1]{fontenc}
\usepackage{times}
\usepackage{graphicx,color}
\usepackage{amsfonts,amsmath,amssymb,amsbsy}
\usepackage[colorlinks=true, linkcolor=blue, citecolor=blue, urlcolor=blue]{hyperref}

\def\D{{\mathcal{D}}}

\let\mathbf=\boldsymbol

\def\emph#1{\textcolor{blue}{#1}}
\begin{document}

\title{Current-Driven Dynamics of Frustrated Skyrmions in a Synthetic Antiferromagnetic Bilayer}

\author{Jing Xia}
\thanks{These authors contributed equally to this work.}
\affiliation{School of Science and Engineering, The Chinese University of Hong Kong, Shenzhen, Guangdong 518172, China}

\author{Xichao Zhang}
\thanks{These authors contributed equally to this work.}
\affiliation{School of Science and Engineering, The Chinese University of Hong Kong, Shenzhen, Guangdong 518172, China}

\author{Motohiko Ezawa}
\email[E-mail:~]{ezawa@ap.t.u-tokyo.ac.jp}
\affiliation{Department of Applied Physics, The University of Tokyo, 7-3-1 Hongo, Tokyo 113-8656, Japan}

\author{Zhipeng Hou}
\affiliation{South China Academy of Advanced Optoelectronics, South China Normal University, Guangzhou 510006, China}
\affiliation{Beijing National Laboratory for Condensed Matter Physics, Institute of Physics, Chinese Academy of Sciences, Beijing 100190, China}

\author{Wenhong Wang}
\affiliation{Beijing National Laboratory for Condensed Matter Physics, Institute of Physics, Chinese Academy of Sciences, Beijing 100190, China}

\author{Xiaoxi Liu}
\affiliation{Department of Electrical and Computer Engineering, Shinshu University, 4-17-1 Wakasato, Nagano 380-8553, Japan}

\author{Yan Zhou}
\email[E-mail:~]{zhouyan@cuhk.edu.cn}
\affiliation{School of Science and Engineering, The Chinese University of Hong Kong, Shenzhen, Guangdong 518172, China}

\begin{abstract}
We report the current-driven dynamics of frustrated skyrmions in an antiferromagnetically exchange coupled bilayer system, where the bilayer skyrmion consists of two monolayer skyrmions with opposite skyrmion numbers $Q$. We show that the in-plane current-driven bilayer skyrmion moves in a straight path, while the out-of-plane current-driven bilayer skyrmion moves in a circular path. It is found that the in-plane current-driven mobility of a bilayer skyrmion is much better than the monolayer one at a large ratio of $\beta/\alpha$, where $\alpha$ and $\beta$ denote the damping parameter and non-adiabatic spin transfer torque strength, respectively. Besides, the out-of-plane current-driven mobility of a bilayer skyrmion is much better than the monolayer one when $\alpha$ is small. We also reveal that one bilayer skyrmion (consisting of monolayer skyrmions with $Q=\pm 2$) can be separated to two bilayer skyrmions (consisting of monolayer skyrmions with $Q=\pm 1$) driven by an out-of-plane current. Our results may be useful for designing skyrmionic devices based on frustrated multilayer magnets.
\end{abstract}

\date{March 27, 2019}
\preprint{}

\maketitle

\section{Introduction}
\label{se:Introduction}

The magnetic skyrmion is a topologically non-trivial object~\cite{Roszler_NATURE2006, Nagaosa_NNANO2013,Wanjun_PHYSREP2017}, which can be regarded as quasi-particles~\cite{Lin_PRB2013} and promises advanced electronic and spintronic applications~\cite{Finocchio_JPD2016,Kang_PIEEE2016,Fert_NATREVMAT2017,Zhou_NSR2018}. For examples, recent studies have suggested that skyrmions can be used as building blocks for racetrack-type memories~\cite{Sampaio_NNANO2013,Tomasello_SREP2014,Guoqiang_NL2017,Muller_NJP2017}, logic computing devices~\cite{Xichao_SREP2015B}, and bio-inspired applications~\cite{Bourianoff_AIP2016,Yangqi_NANO2017,Lisai_NANO2017,Prychynenko_PRAPPL2018}.
The magnetic skyrmion was first experimentally observed in ferromagnetic (FM) materials with Dzyaloshinskii-Moriya (DM) interactions in 2009~\cite{Muhlbauer_SCIENCE2009}. The DM interaction is an essential energy term stabilizing skyrmion textures~\cite{Muhlbauer_SCIENCE2009,Yu_NATURE2010,Du_NCOMMS2015}.
It can also be induced at the interface between the heavy metal and ferromagnet~\cite{Yang_PRL2015}, which promotes recent studies of skyrmions in magnetic bilayer and multilayer structures~\cite{Xichao_NCOMMS2016,Woo_NMATER2016,MoreauLuchaire_NNANO2016,Wanjun_PHYSREP2017,Wanjun_NPHYS2017,Litzius_NPHYS2017,Pollard_NCOMMS2017,Woo_NatElect2018}.

However, some recent studies have demonstrated that skyrmions can be stabilized in frustrated magnets even in the absence of DM interaction~\cite{Leonov_NCOMMS2015,Lin_PRB2016A,Hayami_PRB2016A,Rozsa_PRL2016,Leonov_NCOMMS2017,Yuan_PRB2017,Xichao_NCOMMS2017,Kharkov_PRL2017,Sutcliffe_PRL2017,Hou_AM2017,Liang_NJP2018}, where skyrmions are stabilized by competing exchange interactions~\cite{Leonov_NCOMMS2015,Lin_PRB2016A,Leonov_NCOMMS2017,Xichao_NCOMMS2017}.
Frustrated skyrmions have many unique physical properties compared with skyrmions in DM ferromagnets. For examples, both skyrmions and antiskyrmions can exist in a frustrated magnet as metastable states~\cite{Leonov_NCOMMS2015,Lin_PRB2016A,Leonov_NCOMMS2017,Xichao_NCOMMS2017}. The frustrated skyrmions with skyrmion number of $Q=\pm 1$ can merge and form skyrmions with higher skyrmion number of $Q=\pm 2$~\cite{Xichao_NCOMMS2017}.
Besides, the center-of-mass dynamics of a frustrated skyrmion is coupled to its helicity dynamics~\cite{Leonov_NCOMMS2015,Lin_PRB2016A,Leonov_NCOMMS2017}, which results in the circular motion of a skyrmion~\cite{Lin_PRB2016A,Xichao_NCOMMS2017}.
Therefore, frustrated skyrmions can be used as information carriers, which have multiple degrees of freedom that can be utilized to store information~\cite{Leonov_NCOMMS2015,Lin_PRB2016A,Leonov_NCOMMS2017,Xichao_NCOMMS2017}.

So far, most studies on frustrated skyrmions are focused on the monolayer system, while bilayer and multilayer systems play an important role in developing nanoscale devices~\cite{Parkin_NNANO2015,Xichao_NCOMMS2016,Xichao_NCOMMS2016,Zhang_PRB2016,Woo_NMATER2016,MoreauLuchaire_NNANO2016,Hrabec_NC2017,Tomasello_JPD2017,Koshibae_SREP2017,Wanjun_PHYSREP2017,Wanjun_NPHYS2017,Litzius_NPHYS2017,Pollard_NCOMMS2017,Woo_NatElect2018,Prudnikov_IEEEML2018,Ma_NL2019,Cacilhas_APL2019}.
On the other hand, recent theoretical works have suggested that the monolayer antiferromagnetic (AFM) skyrmion has different dynamics and improved mobility in comparison with the FM one~\cite{Barker_PRL2016,Zhang_SREP2016,Gobel_PRB2017}. For examples, the skyrmion Hall effect can be eliminated in the AFM monolayer, leading to ultra-fast straight motion of a skyrmion that is useful for practical applications~\cite{Barker_PRL2016,Zhang_SREP2016}. However, the dynamic properties of frustrated skyrmions in bilayer and multilayer AFM systems remain elusive.

Here we report the dynamics of frustrated skyrmions with $Q=\pm 1$ and $\pm 2$ in the synthetic AFM bilayer system driven by an in-plane or out-of-plane spin current. We also numerically investigate the current-induced separation of a frustrated skyrmion. We show that the current-driven frustrated skyrmions in the synthetic AFM bilayer system have better mobility than that in the monolayer system.

\section{Methods}
\label{se:Methods}

We consider two FM layers with competing Heisenberg exchange interactions based on the $J_{1}$-$J_{2}$-$J_{3}$ model on a simple square lattice~\cite{Lin_PRB2016A,Xichao_NCOMMS2017}, of which the Hamiltonian is given in Ref.~\onlinecite{SM}.
As shown in Fig.~\ref{FIG1}(a), the two frustrated FM layers are coupled by an AFM interfacial exchange coupling, which can be realized by utilizing the Ruderman-Kittel-Kasuya-Yosida interaction~\cite{Parkin_NNANO2015,Xichao_NCOMMS2016,Zhang_PRB2016,Tomasello_JPD2017,Prudnikov_IEEEML2018,Cacilhas_APL2019}.
The background magnetization direction of the bottom FM layer is assumed to be pointing along the $+z$ direction.
The simulation is performed by using the Object Oriented MicroMagnetic Framework~\cite{OOMMF} with our extension modules for the $J_{1}$-$J_{2}$-$J_{3}$ classical Heisenberg model~\cite{Lin_PRB2016A,Xichao_NCOMMS2017}, where the time-dependent spin dynamics is described by the Landau-Lifshitz-Gilbert (LLG) equation (see Ref.~\onlinecite{SM} for modeling details, parameters, and parameter dependency diagrams).

We consider two geometries for the injection of spin-polarized current.
For the current-in-plane (CIP) geometry, we assume that an in-plane spin current flows along the $+x$ direction in both top and bottom FM layers.
The adiabatic [$\tau_{\text{1}}=u\left(\boldsymbol{m}\times\partial_{x}\boldsymbol{m}\times\boldsymbol{m}\right)$] and non-adiabatic [$\tau_{\text{2}}=-\beta u\left(\boldsymbol{m}\times\partial_{x}\boldsymbol{m}\right)$] spin transfer torque (STT) terms are considered, where $\boldsymbol{m}$ represents the normalized spin, $u=|\frac{\gamma_{0}\hbar}{\mu_{0}e}|\frac{jP}{2M_{\text{S}}}$ is the STT coefficient and $\beta$ is the strength of the non-adiabatic STT torque. $\hbar$ is the reduced Planck constant, $e$ is the electron charge,
$\gamma_{0}$ is the absolute gyromagnetic ratio, $\mu_{0}$ is the vacuum permeability constant,
$j$ is the applied driving current density, $P=0.4$ is the spin polarization rate~\cite{Sampaio_NNANO2013}, and $M_{\text{S}}$ is the saturation magnetization.

For the current-perpendicular-to-plane (CPP) geometry, we assume a heavy-metal substrate layer underneath the bottom FM layer, in which a charge current flows along the $+x$ direction and leads to an out-of-plane spin current propagating into the bottom FM layer due to the spin Hall effect~\cite{Wanjun_NPHYS2017}.
In such a case, the damping-like STT term [$\tau_{\text{d}}=\frac{u}{a}\left(\boldsymbol{m}\times \boldsymbol{p}\times \boldsymbol{m}\right)$] is considered, where $a=0.4$ nm is the thickness of a single FM layer and $\boldsymbol{p}=+\hat{y}$ stands for the unit spin polarization direction.
Note that we ignore the field-like STT term for simplicity as its contribution to the skyrmion dynamics is minor. For the CPP geometry, we set $P=0.1$, which is a typical value of spin Hall angle~\cite{Tomasello_SREP2014,Wanjun_NPHYS2017}.
The skyrmion number in a single FM layer is defined by $Q=-\int d^{2}\boldsymbol{r}\cdot\boldsymbol{m}\cdot\left(\partial_{x}\boldsymbol{m}\times\partial_{y}\boldsymbol{m}\right)/4\pi$.
The internal structure of a skyrmion is described by its helicity number $\eta=[0,2\pi)$ (see Refs.~\onlinecite{Nagaosa_NNANO2013,Koshibae_NCOMMS2016}).

\section{Results and Discussion}
\label{se:Results}

\subsection{Skyrmions driven by an in-plane current}
\label{se:CIP}

\begin{figure}[t]
\centerline{\includegraphics[width=0.46\textwidth]{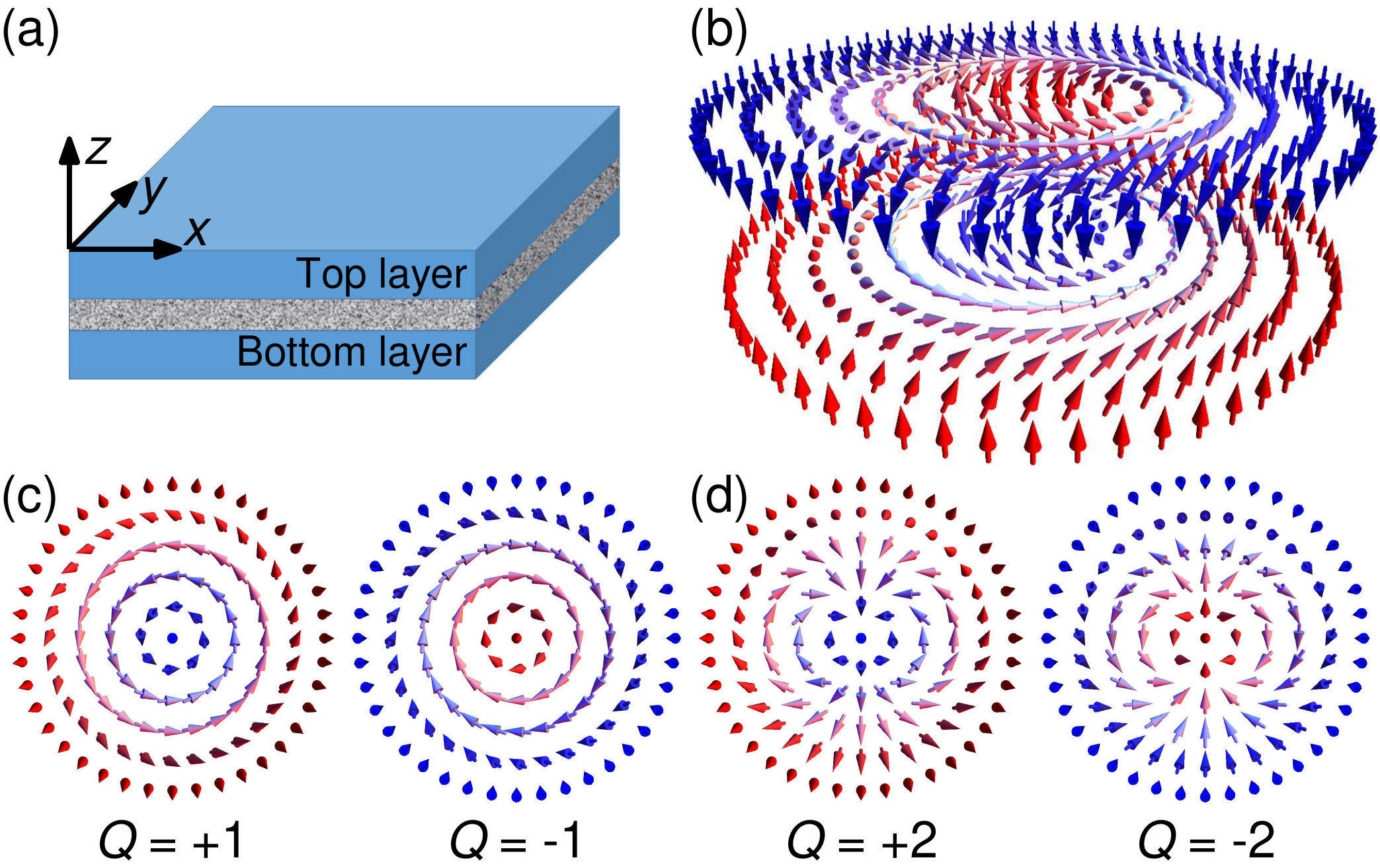}}
\caption{
(a) Schematic of the simulated synthetic AFM bilayer model. The top and bottom FM layers are coupled via an AFM interfacial exchange coupling.
Illustrations of (b) the synthetic AFM bilayer skyrmion, (c) the monolayer skyrmion with $Q=\pm 1$, and (d) the monolayer skyrmion with $Q=\pm 2$.
The arrows represent the magnetization directions. The out-of-plane component of magnetization ($m_z$) is color coded: blue is into the plane, red is out of the plane, white is in-plane.
}
\label{FIG1}
\end{figure}

We first study in-plane current-driven skyrmions in a synthetic AFM bilayer [see Fig.~\ref{FIG1}(a)].
A bilayer skyrmion is first relaxed in the sample, which consists of a top skyrmion with $Q=-1$ and $\eta=3\pi/2$ and a bottom skyrmion with $Q=+1$ and $\eta=\pi/2$ [see Fig.~\ref{FIG1}(b)]. The top and bottom skyrmions are strongly coupled in an AFM manner. In this work, we refer to this bilayer skyrmion as a bilayer skyrmion with $Q=\pm 1$.
The diameter of the relaxed bilayer skyrmion with $Q=\pm 1$ equals $2$ nm.
An in-plane spin current is then injected into the bilayer with the CIP geometry, which drives the bilayer skyrmion into motion.
We also study the skyrmion motion in a single FM monolayer for the purpose of comparison. The monolayer skyrmion structures with $Q=\pm 1$ are illustrated in Fig.~\ref{FIG1}(c). Note that the diameters of the relaxed monolayer and bilayer skyrmions are identical.

\begin{figure}[t]
\centerline{\includegraphics[width=0.46\textwidth]{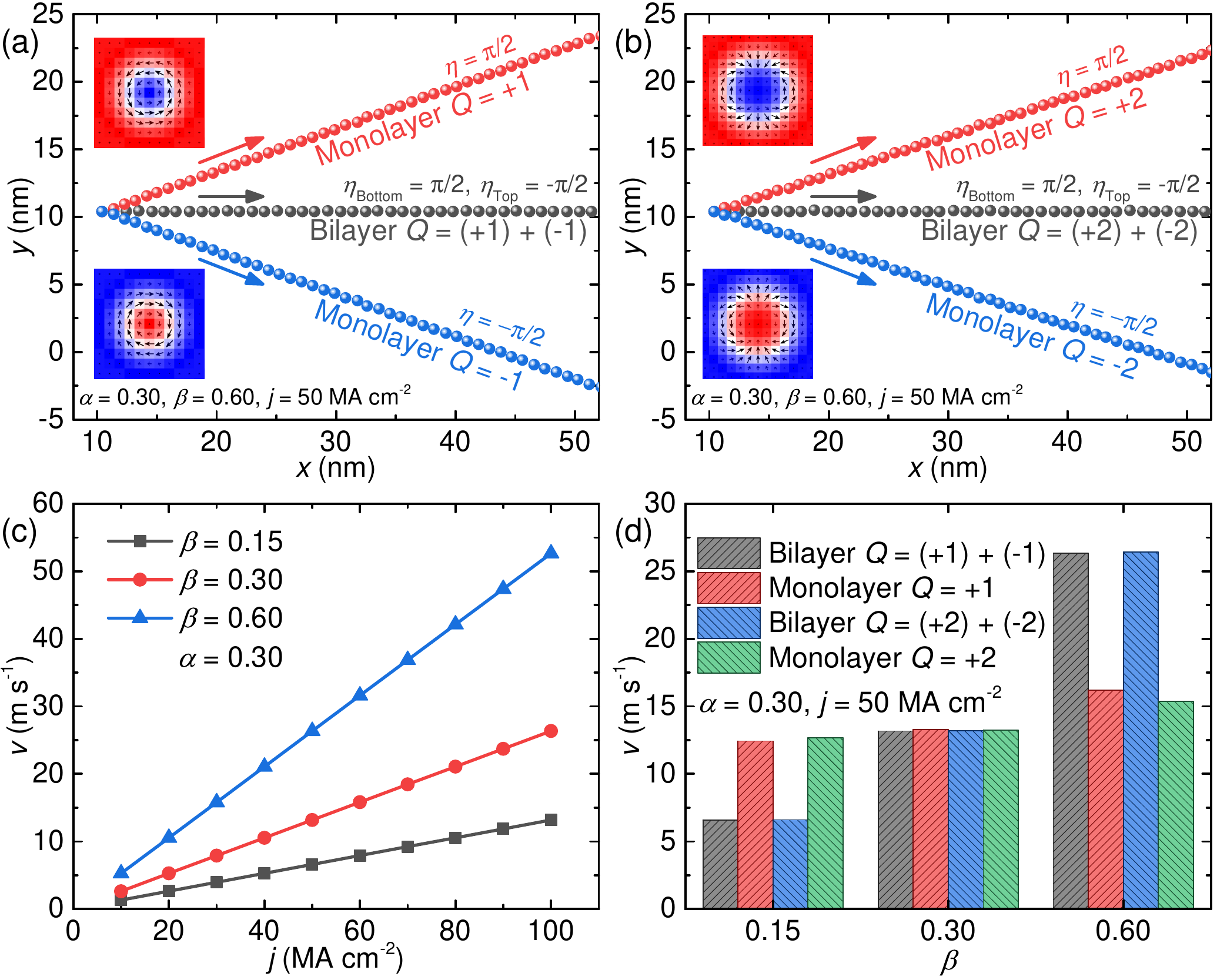}}
\caption{
(a) Trajectories of monolayer and bilayer skyrmions with $Q=\pm 1$ driven by an in-plane current at $\alpha=\beta/2=0.30$.
The arrow denotes the motion direction, and the dot denotes the skyrmion center.
Insets are the top-view snapshots of simulated skyrmions.
(b) Trajectories of monolayer and bilayer skyrmions with $Q=\pm 2$ driven by an in-plane current.
(c) Bilayer skyrmion ($Q=\pm 1$ and $\pm 2$) velocity as a function of in-plane driving current density.
(d) Comparison of the monolayer skyrmion velocity and bilayer skyrmion velocity induced by an in-plane current.
}
\label{FIG2}
\end{figure}

Figure~\ref{FIG2}(a) shows the trajectories of the monolayer and bilayer skyrmions at the damping parameter $\alpha=\beta/2=0.3$. The bilayer skyrmion with $Q=\pm 1$ straightly moves along the $+x$ direction without showing a transverse shift. However, the monolayer skyrmions with $Q=+1$ and $Q=-1$ show transverse shifts toward the $+y$ and $-y$ directions, respectively. The reason is that the monolayer skyrmion experiences a topological Magnus force~\cite{Xichao_NCOMMS2016,Wanjun_NPHYS2017,Litzius_NPHYS2017}, which is perpendicular to the skyrmion velocity vector. Since the directions of Magnus forces acted on monolayer skyrmions with $Q=+1$ and $Q=-1$ are opposite, the bilayer skyrmion consisting of a bottom monolayer skyrmion with $Q=+1$ and a top monolayer skyrmion with $Q=-1$ experiences zero net Magnus force and thus can move in a straight path.
Note that the transverse shift of a skyrmion is referred as the skyrmion Hall effect, which has been observed in experiments~\cite{Wanjun_NPHYS2017,Litzius_NPHYS2017}. The skyrmion Hall effect of the monolayer skyrmion driven by an in-plane current can be eliminated when $\alpha=\beta$ (see Ref.~\onlinecite{Xichao_IEEE2017,SM}), which is difficult to be realized in real materials.
However, the bilayer skyrmion driven by an in-plane current shows no skyrmion Hall effect even when $\alpha\neq\beta$, which is useful for building racetrack-type devices~\cite{Tomasello_SREP2014,Xichao_NCOMMS2016}.

As shown in Fig.~\ref{FIG2}(b), we also simulate the motion of a bilayer skyrmion with $Q=\pm 2$, which consists of a bottom monolayer skyrmion with $Q=+2$ and $\eta=\pi/2$ and a top monolayer skyrmion with $Q=-2$ and $\eta=3\pi/2$ [see Fig.~\ref{FIG1}(d)]. Similarly, the bilayer skyrmion with $Q=\pm 2$ driven by an in-plane current also straightly moves along the $+x$ direction. The monolayer skyrmions with $Q=+2$ and $Q=-2$ show transverse shifts toward the $+y$ and $-y$ directions, respectively.
Note that the diameter of the relaxed bilayer skyrmion with $Q=\pm 2$ equals $2.8$ nm, which is slightly larger than that of the skyrmion with $Q=\pm 1$.

Figure~\ref{FIG2}(c) shows the bilayer skyrmion velocity $v$ as a function of in-plane driving current density $j$ for $\alpha=0.3$ and $\beta=0.15\sim 0.60$. Note that the $v$-$j$ relation of the bilayer skyrmion with $Q=\pm 1$ is identical to that of the bilayer skyrmion with $Q=\pm 2$ at a given $\alpha$ and $\beta$. For both bilayer skyrmions with $Q=\pm 1$ and $Q=\pm 2$, the velocity increases linearly with $j$. At certain values of $\alpha$ and $j$, larger $\beta$ will result in larger velocity.
By comparing the velocities of bilayer and monolayer skyrmions at certain values of $\alpha$ and $j$ [see Fig.~\ref{FIG2}(d)], it is found that bilayer skyrmions move faster than monolayer skyrmions when $\alpha<\beta$, while bilayer skyrmions move slower than monolayer skyrmions when $\alpha>\beta$. When $\alpha=\beta$, the velocities for bilayer and monolayer skyrmions are almost identical.

The center-of-mass dynamics of skyrmions can also be described by the Thiele equation~\cite{Thiele_PRL1973,SM}. For the monolayer skyrmion, the longitudinal (along the $\pm x$ directions) and transverse (along the $\pm y$ directions) velocities induced by an in-plane current are $v_{x}^{m}=u\left(Q^{2}+\alpha\beta\D^{2}\right)/\left(Q^{2}+\alpha^{2}\D^{2}\right)$ and $v_{y}^{m}=u\left[\left(\beta-\alpha\right)Q\D\right]/\left(Q^{2}+\alpha^{2}\D^{2}\right)$, where $\D=\iint\left(\partial_{x}\boldsymbol{m}\right)\cdot\left(\partial_{x}\boldsymbol{m}\right)dxdy/4\pi$ is the dissipative force tensor~\cite{SM}.
Note that we numerically find that $\D$ equals $\sim 1.23$ and $\sim 2.02$ for skyrmions with $Q=\pm 1$ and $Q=\pm 2$ in a single FM layer, respectively.
For the bilayer skyrmion, the longitudinal and transverse velocities induced by an in-plane current are $v_{x}^{b}=u\left(\beta/\alpha\right)$ and $v_{y}^{b}=0$.
Hence, based on the Thiele equation solutions~\cite{SM}, $v_{x}^{m}$ and $v_{y}^{m}$ of the skyrmion with $Q=1$ is $1.03$ and $1.17$ times larger than that of the skyrmion with $Q=2$ at $2\alpha=\beta=0.6$, for example. At the same conditions, $v_{x}^{b}$ is $1.79$ and $1.85$ times larger than $v_{x}^{m}$ of skyrmions with $Q=1$ and $Q=2$. These theoretical results agree well with the simulation results~\cite{SM}.

Several experiments suggest that $\beta$ is considerably larger than $\alpha$ (i.e., $\alpha\ll\beta$)~\cite{Heyne_PRL2010,Pollard_NCOMMS2012}, which means the mobility of bilayer skyrmion is better than that of monolayer skyrmion in real materials.
The monolayer skyrmion speed is proportional to $\beta/\alpha$ in the absence of the skyrmion Hall effect~\cite{Sampaio_NNANO2013}, which is as same as the bilayer skyrmion~\cite{SM}.
However, it should be noted that a merit of the bilayer skyrmion is the enhancement of velocity in large-$\beta$ materials with the CIP geometry, which is promising for future information processing applications with ultrahigh operation speed.

\subsection{Skyrmions driven by an out-of-plane current}
\label{se:CPP}

We also study out-of-plane current-driven skyrmions in a synthetic AFM bilayer. A spin current is perpendicularly injected into the bottom FM layer with the CPP geometry, which drives the skyrmion in the bottom FM layer into motion. The skyrmion in the top FM layer is simultaneously dragged into motion as the bottom and top skyrmions are strongly coupled. We first study the bilayer skyrmion with $Q=\pm 1$ as well as monolayer skyrmions with $Q=+1$ and $Q=-1$.

\begin{figure}[t]
\centerline{\includegraphics[width=0.46\textwidth]{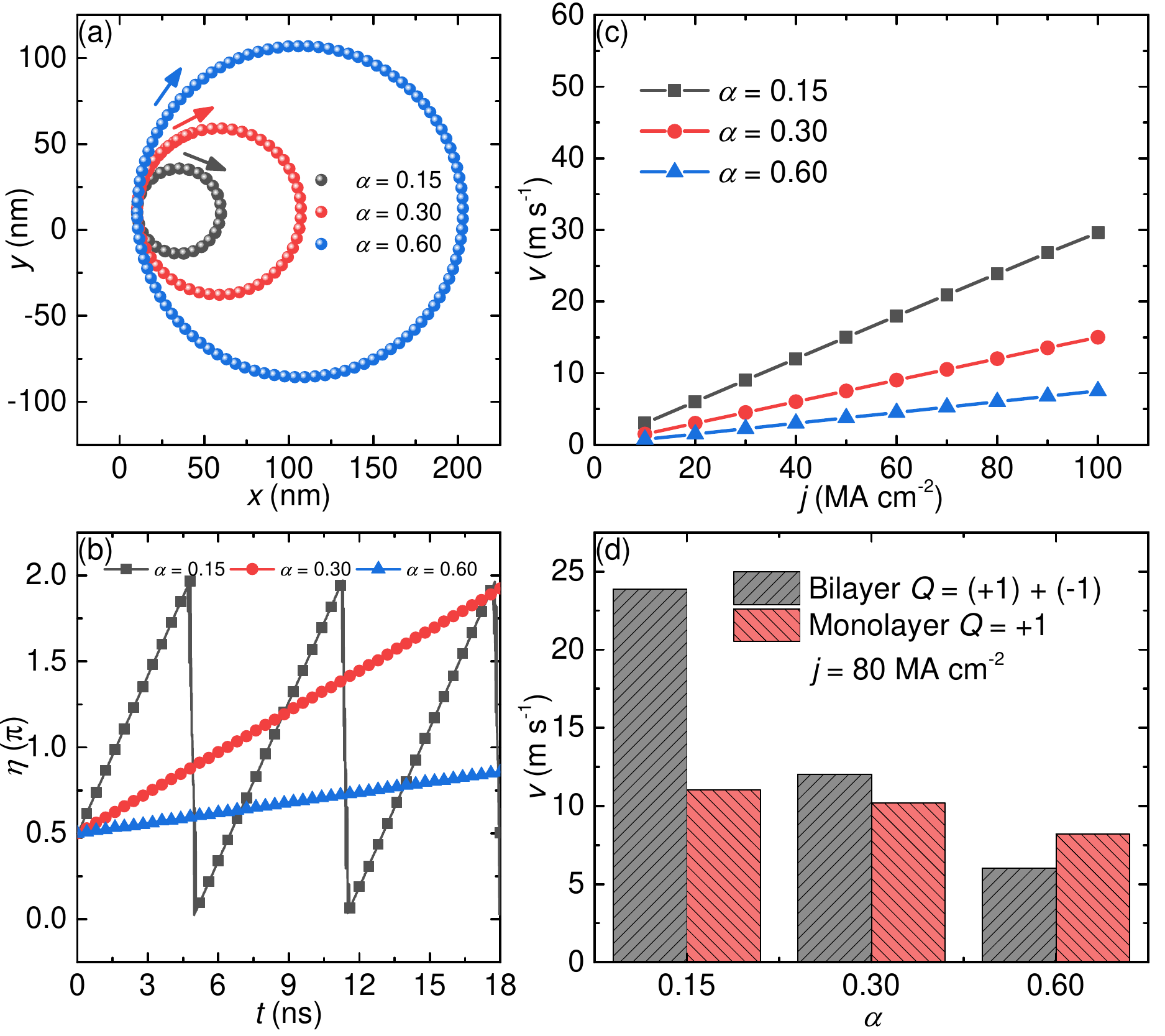}}
\caption{
(a) Trajectories of bilayer skyrmions with $Q=\pm 1$ driven by an out-of-plane current.
The arrow denotes the motion direction, and the dot denotes the skyrmion center.
(b) Helicity as a function of time during the motion of bilayer skyrmions with $Q=\pm 1$. Note that the helicity of the skyrmion in bottom FM layer is shown.
(c) Bilayer skyrmion ($Q=\pm 1$) velocity as a function of out-of-plane driving current density.
(d) Comparison of the monolayer skyrmion velocity and bilayer skyrmion velocity induced by an out-of-plane current.
}
\label{FIG3}
\end{figure}

The dynamics of a bilayer skyrmion with $Q=\pm 1$ [see Fig.~\ref{FIG3}(a)] and its monolayer counterparts (see Ref.~\onlinecite{SM}) driven by an out-of-plane current are in stark contrast to the case driven by an in-plane current.
It is found that the bilayer skyrmion with $Q=\pm 1$ and monolayer skyrmions with $Q=+1$ and $Q=-1$ driven by an out-of-plane current move in a circular path. The bilayer skyrmion with $Q=\pm 1$ moves in the clockwise direction, while the monolayer skyrmions with $Q=+1$ and $Q=-1$ move in the counterclockwise and clockwise directions, respectively.
As pointed out in Ref.~\onlinecite{Lin_PRB2016A}, the motion direction depends on the skyrmion helicity.
For both the bilayer and monolayer skyrmions, the diameters of their circular trajectories increase with increasing $\alpha$.
Under the same $j$ and $\alpha$, the trajectory diameter of the bilayer skyrmion is significantly larger than that of the monolayer skyrmion, which means the frustrated skyrmion can be delivered farther in the synthetic AFM bilayer.

On the other hand, the helicity numbers $\eta$ of bilayer skyrmion with $Q=\pm 1$ [see Fig.~\ref{FIG3}(b)] and monolayer skyrmions with $Q=+1$ and $Q=-1$ (see Ref.~\onlinecite{SM}) driven by an out-of-plane current are coupled to their center-of-mass dynamics. Namely, $\eta$ changes linearly with time during the skyrmion motion.
Such a phenomenon has been reported for monolayer skyrmions~\cite{Lin_PRB2016A}, and we show that it also happens for bilayer skyrmions with $Q=\pm 1$. As pointed out in Ref.~\onlinecite{Lin_PRB2016A}, the frustrated monolayer skyrmion has a translational mode and a rotational mode, which can be excited and hybridized by the damping-like STT. Namely, when the skyrmion in the bottom FM layer is driven by the damping-like STT, it moves along a circle with rotating helicity. At the same time, it drives the skyrmion in the top FM layer into circular motion with rotating helicity as the bottom and top skyrmions are coupled.

Figure~\ref{FIG3}(c) shows the velocity of the bilayer skyrmion with $Q=\pm 1$ as a function of out-of-plane driving current density. At a given $\alpha$, the velocity of bilayer skyrmion with $Q=\pm 1$ increases linearly with $j$. At a given $j$, the velocity of bilayer skyrmion with $Q=\pm 1$ is inversely proportional to $\alpha$. By comparing Fig.~\ref{FIG2}(c) and Fig.~\ref{FIG3}(c) at same $j$ and $\alpha$, it is found that the bilayer skyrmion with $Q=\pm 1$ driven by an in-plane current can move much faster than the one driven by an out-of-plane current if $\beta$ is large. For example, when $j=100$ MA cm$^{-2}$ and $\alpha=0.3$, the bilayer skyrmion with $Q=\pm 1$ driven by an in-plane current can reach a velocity of $v=53$ m s$^{-1}$ at $\beta=0.6$, while the one driven by an out-of-plane current only reaches a velocity of $v=15$ m s$^{-1}$.
However, it should be noted that the driving force provide by the in-plane current is larger than that provided by the out-of-plane current at the same $j$ in this work, as the CIP spin-polarization rate is four times larger than the CPP spin Hall angle. The bilayer skyrmion driven by an out-of-plane current can move faster when the value of the spin Hall angle is increased.

Besides, by comparing the velocities of bilayer skyrmion with $Q=\pm 1$ and monolayer skyrmions with $Q=+1$ and $Q=-1$ at a given $j$ [see Fig.~\ref{FIG3}(d)], it is found that the bilayer skyrmion velocity remarkably decreases with increasing $\alpha$, while the monolayer skyrmion velocity slightly decreases with increasing $\alpha$.
The bilayer skyrmion velocity is much larger than monolayer skyrmion velocity at small $\alpha$ (e.g., the bilayer skyrmion is $2.2$ times faster than the monolayer one at $\alpha=0.15$), while it is slightly smaller than monolayer skyrmion velocity at large $\alpha$ (e.g., $\alpha=0.60$).

The circular motion and helicity dynamics of the monolayer and bilayer skyrmions with $Q=\pm 1$ induced by an out-of-plane current can also be analyzed by the Thiele equation~\cite{Thiele_PRL1973,SM}.
First, for the monolayer skyrmion with $Q=\pm 1$~\cite{SM}, the velocity can be expressed as $\boldsymbol{v}^{m}=\boldsymbol{R}\left(-\eta\right)\boldsymbol{v}_{\eta=0}^{m}$, where $\boldsymbol{R}\left(-\eta\right)$ is a counterclockwise rotation matrix determined by $\eta$~\cite{SM}, and $\boldsymbol{v}_{\eta=0}^{m}$ is the velocity at $\eta=0$. The transverse and longitudinal components of $\boldsymbol{v}_{\eta=0}^{m}$ are found to be $v_{x}^{m}=(uI\alpha\D)/(\alpha^{2}\D^{2}+Q^{2})$ and $v_{y}^{m}=(uIQ)/(\alpha^{2}\D^{2}+Q^{2})$, where $I=\pi r_{sk}/4a$ with $r_{sk}$ being the skyrmion radius. Namely, the magnitude of the skyrmion velocity $v^{m}=uI/\sqrt{\alpha^{2}\D^{2}+Q^{2}}$, which is independent of $\eta$.

Second, for the bilayer skyrmion with $Q=\pm 1$~\cite{SM}, the velocity can also be expressed as $\boldsymbol{v}^{b}=\boldsymbol{R}\left(-\eta\right)\boldsymbol{v}_{\eta=0}^{b}$. The transverse and longitudinal components of $\boldsymbol{v}_{\eta=0}^{b}$ are found to be $v_{x}^{b}=uI/2\alpha\D$ and $v_{y}^{b}=0$. Similarly, the magnitude of the skyrmion velocity $v^{b}=uI/2\alpha\D$.
It can be seen that the monolayer skyrmion speed $v^{m}$ decreases from $uI$ to $uI/1.56$ as $\alpha$ increases from $0$ to $1$. The bilayer skyrmion speed $v^{b}$ is also inversely proportional to $\alpha$. It is found that the bilayer skyrmion moves faster than the monolayer one when $\alpha<0.47$. These theoretical results are consistent with the simulation results [see Fig.~\ref{FIG3}(d)].

\begin{figure}[t]
\centerline{\includegraphics[width=0.46\textwidth]{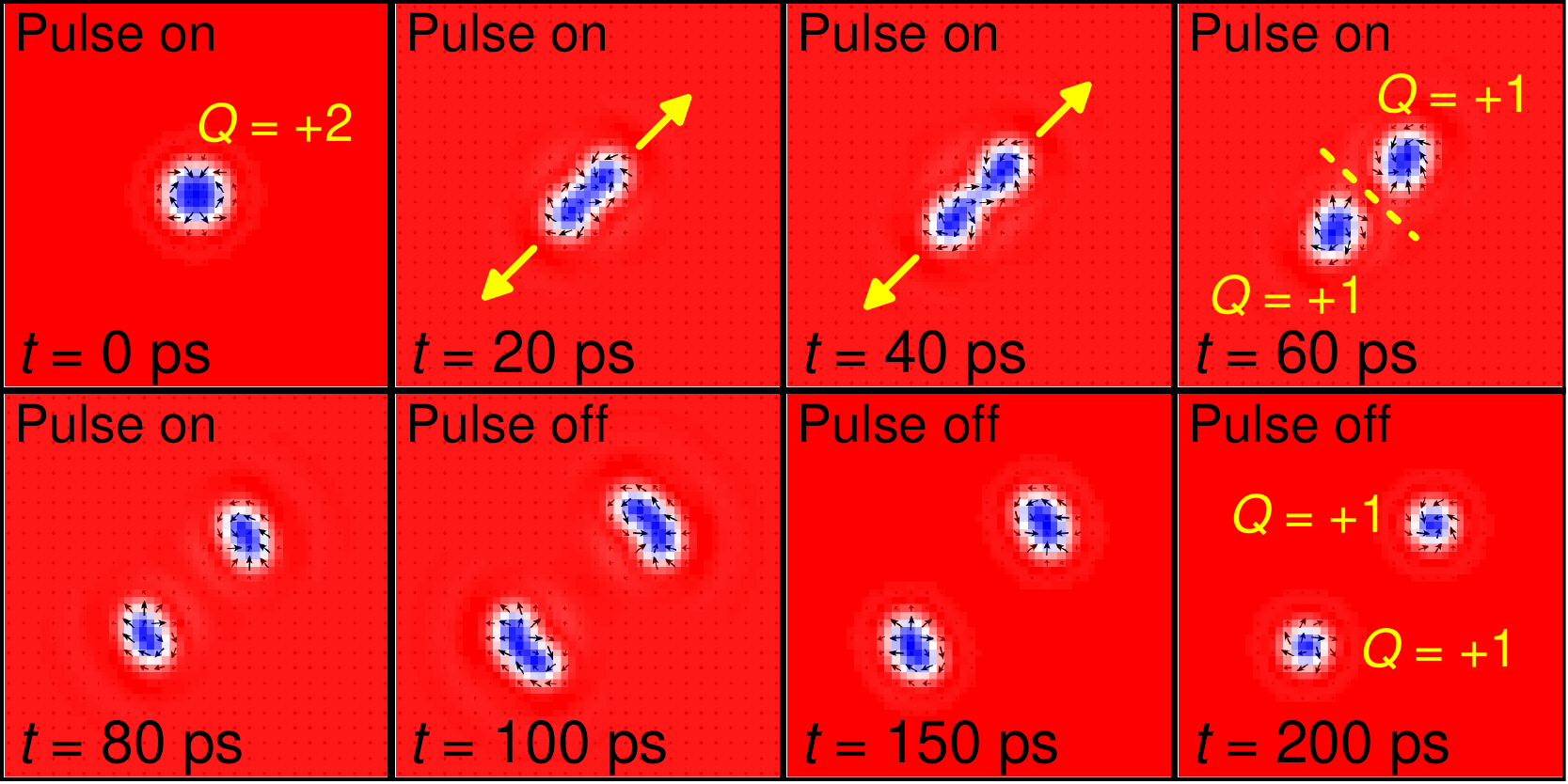}}
\caption{
Out-of-plane current-induced separation of one bilayer skyrmion with $Q=\pm 2$ to two bilayer skyrmions with $Q=\pm 1$.
Snapshots are top views of the bottom FM layer. Here, a current pulse of $j=400$ MA cm$^{-2}$ is applied for $100$ ps to force the separation.
}
\label{FIG4}
\end{figure}

We continue to numerically study the bilayer skyrmion with $Q=\pm 2$ and monolayer skyrmions with $Q=+2$ and $Q=-2$ driven by an out-of-plane current.
The bilayer skyrmion with $Q=\pm 2$ does not move toward a certain direction when an out-of-plane current is applied (see Fig.~\ref{FIG4}).
Instead, it is elongated and then separated to two bilayer skyrmions with $Q=\pm 1$.
The reason is that the bilayer skyrmion with $Q=\pm 2$ is topologically equal to a combination of two bilayer skyrmions with $Q=\pm 1$ and opposite $\eta$. Such two bilayer skyrmions with $Q=\pm 1$ tend to move along opposite directions upon the application of an out-of-plane current, as indicated by yellow arrows in Fig.~\ref{FIG4}.

Similarly, when an out-of-plane current is applied, the monolayer skyrmion with $Q=+2$ is separated to two monolayer skyrmions with $Q=+1$ and the monolayer skyrmion with $Q=-2$ is separated to two monolayer skyrmions with $Q=-1$. The most obvious difference between the out-of-plane current-induced separation of a bilayer skyrmion with $Q=\pm 2$ and the separation of a monolayer skyrmion with $Q=+2$ or $Q=-2$ is that the separation of a monolayer skyrmion is accompanied with an obvious clockwise or counterclockwise rotation of the skyrmion structure.
The reason is that the out-of-plane current-induced circular motion diameter of the monolayer skyrmion with $Q=+1$ or $Q=-1$ is much smaller than that of the bilayer skyrmion with $Q=\pm 1$. Hence, the separation of the monolayer skyrmion with $Q=\pm 2$ shows very obvious rotation caused by the circular motion of monolayer skyrmions with $Q=+1$ and $Q=-1$.
Note that if an out-of-plane current is applied after the separation of the bilayer skyrmion with $Q=\pm 2$, it will also drive the bilayer skyrmions with $Q=\pm 1$ into motion along a large circle.

\section{Conclusion}
\label{se:Conclusion}

In conclusion, we have studied the current-driven dynamics of frustrated bilayer skyrmions in a synthetic AFM bilayer.
It is found that the bilayer skyrmion with $Q=\pm 1$ or $Q=\pm 2$ moves in a straight path driven by an in-plane current, while the bilayer skyrmion with $Q=\pm 1$ moves in a circular path driven by an out-of-plane current.
Therefore, the bilayer skyrmion driven by an in-plane current is promising for building racetrack-type applications, as it shows no skyrmion Hall effect and can be delivered in a straight line even when $\alpha\neq\beta$.
Especially, as the synthetic AFM bilayer skyrmions with $Q=\pm 1$ and $Q=\pm 2$ show the same velocity-current relation (i.e., $v_{x}^{b}=u\beta/\alpha$) and no skyrmion Hall effect (i.e., $v_{y}^{b}=0$), therefore, it is possible to build a racetrack-type memory based on the in-line motion of skyrmions with $Q=\pm 1$ and $Q=\pm 2$, where the elementary binary information digits ``1'' and ``0'' are encoded and carried by skyrmions with $Q=\pm 1$ and $Q=\pm 2$, respectively.

The elementary binary information digits in traditional monolayer skyrmion-based racetrack-type memory designs are encoded by the skyrmion and ferromagnetic background. However, because the spacing between moving skyrmions is easy to be changed due to the pinning effect~\cite{Lin_PRB2013,Reichhardt_NJP2016}, the information stored in the racetrack may be distorted or destroyed. Obviously, such a problem can be reliably avoided by using the synthetic AFM bilayer skyrmions with different $Q$, as the spacing between bilayer skyrmions with $Q=\pm 1$ and $Q=\pm 2$ is irrelevant to the information encoding.
Besides, at same $j$, the mobility of the bilayer skyrmion driven by an in-plane current is better than that of the monolayer skyrmion at large $\beta/\alpha$.

On the other hand, the monolayer and bilayer skyrmions with $Q=\pm 1$ can be driven into circular motion by an out-of-plane current.
First, it is noteworthy that the maximum speed of the monolayer skyrmion with $Q=\pm 1$ induced by a given $j$ approaches $uI$ as $\alpha\rightarrow 0$. However, there is no maximum speed limit for the bilayer skyrmion with $Q=\pm 1$ induced by a given $j$, because its speed is determined by $uI/2\alpha\D$. Namely, the bilayer skyrmion can move in an extremely fast manner when $\alpha$ is very small.

Form an application point of view, the circular motion of skyrmions can be harnessed for building skyrmion-based oscillators~\cite{Senfu_NJP2015,LiuQF_PRAP2018}, which generate high-frequency microwave signals.
As the helicity of the skyrmion with $Q=\pm 1$ driven by the out-of-plane current is coupled to its translational motion, which means the helicity can be controlled by manipulating the motion of the skyrmion. With this feature, it is also possible to build skyrmion helicity-based multi-state memory devices~\cite{Xichao_NCOMMS2017}, where different skyrmion helicity numbers stand for different elementary information digits. For example, the Bloch-type skyrmions with $\eta=\pi/2$ and $\eta=3\pi/2$ can be used to carry different information.

In addition, the out-of-plane current can lead to the separation of a bilayer skyrmion with $Q=\pm 2$ to two bilayer skyrmions with $Q=\pm 1$. The separation of a bilayer skyrmion with $Q=\pm 2$ does not show the rotation of skyrmion structure, while the monolayer skyrmion with $Q=+2$ or $Q=-2$ is rotating until it is separated to two monolayer skyrmions with $Q=+1$ or $Q=-1$.
The forced separation of skyrmions with higher $Q$ (i.e., excited states) to skyrmions with lower $Q$ (i.e., ground states) could be an important operation in future multi-state memory devices, which are based on the manipulation of $Q$.
Our results are useful for understanding bilayer skyrmion physics in frustrated magnets and could provide guidelines for building skyrmion-based devices.

\begin{acknowledgments}
X.Z. acknowledges the support by the Presidential Postdoctoral Fellowship of the Chinese University of Hong Kong, Shenzhen (CUHKSZ).
M.E. acknowledges the support by the Grants-in-Aid for Scientific Research from JSPS KAKENHI (Grant Nos. JP18H03676, JP17K05490 and JP15H05854), and the support by CREST, JST (Grant Nos. JPMJCR1874 and JPMJCR16F1).
Z.H. and W.W. acknowledge the support by the National Key R\&D Program of China (Grant Nos. 2017YFA0303202 and 2017YFA206303), the National Natural Science Foundation of China (Grant Nos. 11604148 and 11874410), and the Key Research Program of the Chinese Academy of Sciences (Grant No. KJZD-SW-M01).
X.L. acknowledges the support by the Grants-in-Aid for Scientific Research from JSPS KAKENHI (Grant Nos. JP17K19074, 26600041 and 22360122).
Y.Z. acknowledges the support by the President's Fund of CUHKSZ, the National Natural Science Foundation of China (Grant No. 11574137), and Shenzhen Fundamental Research Fund (Grant Nos. JCYJ20160331164412545 and JCYJ20170410171958839).
\end{acknowledgments}




\begin{thebibliography}{99}

\bibitem{Roszler_NATURE2006} U.~K. R{\"o}{\ss}ler, A.~N. Bogdanov, and C. Pfleiderer, Spontaneous skyrmion ground states in magnetic metals, Nature \textbf{442}, 797 (2006).

\bibitem{Nagaosa_NNANO2013} N. Nagaosa and Y. Tokura, Topological properties and dynamics of magnetic skyrmions, Nat. Nanotech. \textbf{8}, 899 (2013).

\bibitem{Wanjun_PHYSREP2017} W. Jiang, G. Chen, K. Liu, J. Zang, S.~G. Velthuiste, and A. Hoffmann, Skyrmions in magnetic multilayers, Phys. Rep. \textbf{704}, 1 (2017).

\bibitem{Lin_PRB2013} S.-Z. Lin, C. Reichhardt, C. D. Batista, and A. Saxena, Particle model for skyrmions in metallic chiral magnets: dynamics, pinning, and creep, Phys. Rev. B \textbf{87}, 214419 (2013).

\bibitem{Finocchio_JPD2016} G. Finocchio, F. B{\"u}ttner, R. Tomasello, M. Carpentieri and M. Kl{\"a}ui, Magnetic skyrmions: from fundamental to applications, J. Phys. D: Appl. Phys. \textbf{49}, 423001 (2016).

\bibitem{Kang_PIEEE2016} W. Kang, Y. Huang, X. Zhang, Y. Zhou, and W. Zhao, Skyrmion-electronics: an overview and outlook, Proc. IEEE \textbf{104}, 2040 (2016).

\bibitem{Fert_NATREVMAT2017} A. Fert, N. Reyren, and V. Cros, Magnetic skyrmions: advances in physics and potential applications, Nat. Rev. Mater. \textbf{2}, 17031 (2017).

\bibitem{Zhou_NSR2018} Y. Zhou, Magnetic skyrmions: intriguing physics and new spintronic device concepts. Natl. Sci. Rev. \textbf{6}, 1 (2018).

\bibitem{Sampaio_NNANO2013} J. Sampaio, V. Cros, S. Rohart, A. Thiaville, and A. Fert, Nucleation, stability and current-induced motion of isolated magnetic skyrmions in nanostructures, Nat. Nanotech. \textbf{8}, 839 (2013).

\bibitem{Tomasello_SREP2014} R. Tomasello, E. Martinez, R. Zivieri, L. Torres, M. Carpentieri, and G. Finocchio, A strategy for the design of skyrmion racetrack memories, Sci. Rep. \textbf{4}, 6784 (2014).

\bibitem{Guoqiang_NL2017} G. Yu, P. Upadhyaya, Q. Shao, H. Wu, G. Yin, X. Li, C. He, W. Jiang, X. Han, P.~K. Amiri, and K.~L. Wang, Room-temperature skyrmion shift device for memory application, Nano Lett. \textbf{17}, 261 (2017).

\bibitem{Muller_NJP2017} J. M{\"u}ller, Magnetic skyrmions on a two-lane racetrack, New J. Phys. \textbf{19}, 025002 (2017).

\bibitem{Xichao_SREP2015B} X. Zhang, M. Ezawa, and Y. Zhou, Magnetic skyrmion logic gates: conversion, duplication and merging of skyrmions, Sci. Rep. \textbf{5}, 9400 (2015).

\bibitem{Yangqi_NANO2017} Y. Huang, W. Kang, X. Zhang, Y. Zhou, and W. Zhao, Magnetic skyrmion-based synaptic devices, Nanotechnology \textbf{28}, 08LT02 (2017).

\bibitem{Lisai_NANO2017} S. Li, W. Kang, Y. Huang, X. Zhang, Y. Zhou, and W. Zhao, Magnetic skyrmion-based artificial neuron device, Nanotechnology \textbf{28}, 31LT01 (2017).

\bibitem{Prychynenko_PRAPPL2018} D. Prychynenko, M. Sitte, K. Litzius, B. Kr{\"u}ger, G. Bourianoff, M. Kl{\"a}ui, J. Sinova, and K. Everschor-Sitte, Magnetic skyrmion as a nonlinear resistive element: a potential building block for reservoir computing, Phys. Rev. Appl. \textbf{9}, 014034 (2018).

\bibitem{Bourianoff_AIP2016} G. Bourianoff, D. Pinna, M. Sitte, and K. Everschor-Sitte, Potential implementation of reservoir computing models based on magnetic skyrmions, AIP Adv. \textbf{8}, 055602 (2018).

\bibitem{Muhlbauer_SCIENCE2009} S. M{\"u}hlbauer, B. Binz, F. Jonietz, C. Pfleiderer, A. Rosch, A. Neubauer, R. Georgii, and P. B{\"o}ni, Skyrmion lattice in a chiral magnet, Science \textbf{323}, 915 (2009).

\bibitem{Yu_NATURE2010} X.~Z. Yu, Y. Onose, N. Kanazawa, J.~H. Park, J.~H. Han, Y. Matsui, N. Nagaosa, and Y. Tokura, Real-space observation of a two-dimensional skyrmion crystal, Nature \textbf{465}, 901 (2010).

\bibitem{Du_NCOMMS2015} H. Du, R. Che, L. Kong, X. Zhao, C. Jin, C. Wang, J. Yang, W. Ning, R. Li, C. Jin, X. Chen, J. Zang, Y. Zhan, and M. Tian, Edge-mediated skyrmion chain and its collective dynamics in a confined geometry, Nat. Commun. \textbf{6}, 8504 (2015).

\bibitem{Yang_PRL2015} H. Yang, A. Thiaville, S. Rohart, A. Fert, and M. Chshiev, Anatomy of Dzyaloshinskii-Moriya interaction at Co/Pt interfaces, Phys. Rev. Lett. \textbf{115}, 267210 (2015).

\bibitem{Woo_NMATER2016} S. Woo, K. Litzius, B. Kruger, M.-Y. Im, L. Caretta, K. Richter, M. Mann, A. Krone, R.~M. Reeve, M. Weigand, P. Agrawal, I. Lemesh, M.-A. Mawass, P. Fischer, M. Klaui, and G.~S.~D. Beach, Observation of room-temperature magnetic skyrmions and their current-driven dynamics in ultrathin metallic ferromagnets, Nat. Mater. \textbf{15}, 501 (2016).

\bibitem{MoreauLuchaire_NNANO2016} C. Moreau-Luchaire, C. Moutafis, N. Reyren, J. Sampaio, C.~A.~F. Vaz, N. Van~Horne, K. Bouzehouane, K. Garcia, C. Deranlot, P. Warnicke, P. Wohlh\"{u}ter, J.-M. George, M. Weigand, J. Raabe, V. Cros, and A. Fert, Additive interfacial chiral interaction in multilayers for stabilization of small individual skyrmions at room temperature, Nat. Nanotech. \textbf{11}, 444 (2016).

\bibitem{Pollard_NCOMMS2017} S.~D. Pollard, J.~A. Garlow, J. Yu, Z. Wang, Y. Zhu, and H. Yang, Observation of stable Néel skyrmions in cobalt/palladium multilayers with Lorentz transmission electron microscopy, Nat. Commun. \textbf{8}, 14761 (2017).

\bibitem{Woo_NatElect2018} S. Woo, K. M. Song, X. Zhang, M. Ezawa, Y. Zhou, X. Liu, M. Weigand, S. Finizio, J. Raabe, M.-C. Park, K.-Y. Lee, J. W. Choi, B.-C. Min, H. C. Koo, and J. Chang, Deterministic creation and deletion of a single magnetic skyrmion observed by direct time-resolved X-ray microscopy, Nat. Electron. \textbf{1}, 288 (2018).

\bibitem{Xichao_NCOMMS2016} X. Zhang, Y. Zhou, and M. Ezawa, Magnetic bilayer-skyrmions without skyrmion Hall effect, Nat. Commun. \textbf{7}, 10293 (2016).

\bibitem{Wanjun_NPHYS2017} W. Jiang, X. Zhang, G. Yu, W. Zhang, X. Wang, M. Benjamin~Jungfleisch, J.~E. Pearson, X. Cheng, O. Heinonen, K.~L. Wang, Y. Zhou, A. Hoffmann, and S.~G.~E. Velthuiste, Direct observation of the skyrmion Hall effect, Nat. Phys. \textbf{13}, 162 (2017).

\bibitem{Litzius_NPHYS2017} K. Litzius, I. Lemesh, B. Kruger, P. Bassirian, L. Caretta, K. Richter, F. Buttner, K. Sato, O.~A. Tretiakov, J. Forster, R.~M. Reeve, M. Weigand, I. Bykova, H. Stoll, G. Schutz, G.~S.~D. Beach, and M. Klaui, Skyrmion Hall effect revealed by direct time-resolved X-ray microscopy, Nat. Phys. \textbf{13}, 170 (2017).

\bibitem{Leonov_NCOMMS2015} A.~O. Leonov and M. Mostovoy, Multiply periodic states and isolated skyrmions in an anisotropic frustrated magnet, Nat. Commun. \textbf{6}, 8275 (2015).

\bibitem{Lin_PRB2016A} S.-Z. Lin and S. Hayami, Ginzburg-Landau theory for skyrmions in inversion-symmetric magnets with competing interactions, Phys. Rev. B \textbf{93}, 064430 (2016).

\bibitem{Hayami_PRB2016A} S. Hayami, S.-Z. Lin, and C.~D. Batista, Bubble and skyrmion crystals in frustrated magnets with easy-axis anisotropy, Phys. Rev. B \textbf{93}, 184413 (2016).

\bibitem{Rozsa_PRL2016} L. R{\'o}zsa, A. De{\'a}k, E. Simon, R. Yanes, L. Udvardi, L. Szunyogh, and U. Nowak, Skyrmions with attractive interactions in an ultrathin magnetic film, Phys. Rev. Lett. \textbf{117}, 157205 (2016).

\bibitem{Leonov_NCOMMS2017} A.~O. Leonov and M. Mostovoy, Edge states and skyrmion dynamics in nanostripes of frustrated magnets, Nat. Commun. \textbf{8}, 14394 (2017).

\bibitem{Xichao_NCOMMS2017} X. Zhang, J. Xia, Y. Zhou, X. Liu, H. Zhang, and M. Ezawa, Skyrmion dynamics in a frustrated ferromagnetic film and current-induced helicity locking-unlocking transition, Nat. Commun. \textbf{8}, 1717 (2017).

\bibitem{Yuan_PRB2017} H.~Y. Yuan, O. Gomonay, and M. Kl{\"a}ui, Skyrmions and multisublattice helical states in a frustrated chiral magnet, Phys. Rev. B \textbf{96}, 134415 (2017).

\bibitem{Kharkov_PRL2017} Y.~A. Kharkov, O.~P. Sushkov, and M. Mostovoy, Bound states of skyrmions, and merons near the Lifshitz point, Phys. Rev. Lett. \textbf{119}, 207201 (2017).

\bibitem{Hou_AM2017} Z. Hou, W. Ren, B. Ding, G. Xu, Y. Wang, B. Yang, Q. Zhang, Y. Zhang, E. Liu, F. Xu, W. Wang, G. Wu, X. Zhang, B. Shen, and Z. Zhang, Observation of various and spontaneous magnetic skyrmionic bubbles at room temperature in a frustrated kagome magnet with uniaxial magnetic anisotropy, Adv. Mater. \textbf{29}, 1701144 (2017).

\bibitem{Sutcliffe_PRL2017} P. Sutcliffe, Skyrmion knots in frustrated magnets, Phys. Rev. Lett. \textbf{118}, 247203 (2017).

\bibitem{Liang_NJP2018} J.~J. Liang, J.~H. Yu, J. Chen, M.~H. Qin, M. Zeng, X.~B. Lu, X.~S. Gao, and J. Liu, Magnetic field gradient driven dynamics of isolated skyrmions and antiskyrmions in frustrated magnets, New J Phys. \textbf{20}, 053037 (2018).

\bibitem{Parkin_NNANO2015} S.-H. Yang, K.-S. Ryu, and S. Parkin, Domain-wall velocities of up to 750 m s$^{-1}$ driven by exchange-coupling torque in synthetic antiferromagnets, Nat. Nanotech. \textbf{10}, 221 (2015).

\bibitem{Prudnikov_IEEEML2018} A. Prudnikov, M. Li, M. D. Graef, and V. Sokalski, Simultaneous control of interlayer exchange coupling and the interfacial Dzyaloshinskii-Moriya interaction in Ru-based synthetic antiferromagnets, IEEE Magn. Lett. \textbf{10}, 6100304 (2019).

\bibitem{Zhang_PRB2016} X. Zhang, M. Ezawa, and Y. Zhou, Thermally stable magnetic skyrmions in multilayer synthetic antiferromagnetic racetracks, Phys. Rev. B \textbf{94}, 064406 (2016).

\bibitem{Tomasello_JPD2017} R. Tomasello, V. Puliafito, E. Martinez, A. Manchon, M. Ricci, M. Carpentieri, and G. Finocchio, Performance of synthetic antiferromagnetic racetrack memory: domain wall versus skyrmion, J. Phys. D: Appl. Phys. \textbf{50}, 325302 (2017).

\bibitem{Koshibae_SREP2017} W. Koshibae and N. Nagaosa, Theory of skyrmions in bilayer systems, Sci. Rep. \textbf{7}, 42645 (2017).

\bibitem{Ma_NL2019} C. Ma, X. Zhang, J. Xia, M. Ezawa, W. Jiang, T. Ono, S. N. Piramanayagam, A. Morisako, Y. Zhou, and X. Liu, Electric field-induced creation and directional motion of domain walls and skyrmion bubbles, Nano Lett. \textbf{19}, 353 (2019).

\bibitem{Hrabec_NC2017} A. Hrabec, J. Sampaio, M. Belmeguenai, I. Gross, R. Weil, S. M. Ch{\'e}rif, A. Stashkevich, V. Jacques, A. Thiaville, and S. Rohart, Current-induced skyrmion generation and dynamics in symmetric bilayers, Nat. Commun. \textbf{8}, 15765 (2017).

\bibitem{Cacilhas_APL2019} R. Cacilhas, V. L. Carvalho-Santos, S. Vojkovic, E. B. Carvalho, A. R. Pereira, D. Altbir, and {\'A}. S. N{\'u}{\~n}ez, Coupling of skyrmions mediated by the RKKY interaction, Appl. Phys. Lett. \textbf{113}, 212406 (2018).

\bibitem{Barker_PRL2016} J. Barker and O. A. Tretiakov, Static and dynamical properties of antiferromagnetic skyrmions in the presence of applied current and temperature, Phys. Rev. Lett. \textbf{116}, 147203 (2016).

\bibitem{Zhang_SREP2016} X. Zhang, Y. Zhou, and M. Ezawa, Antiferromagnetic skyrmion: stability, creation and manipulation, Sci. Rep. \textbf{6}, 24795 (2016).

\bibitem{Gobel_PRB2017} B. G{\"o}bel, A. Mook, J. Henk, and I. Mertig, Antiferromagnetic skyrmion crystals: generation, topological Hall, and topological spin Hall effect, Phys. Rev. B \textbf{96}, 060406(R) (2017).

\bibitem{SM} See Supplemental Material at [URL] for the simulation modeling details and more information regarding the current-driven motion and separation of monolayer skyrmions, parameter dependency diagrams, and Thiele equations for monolayer and bilayer skyrmions.

\bibitem{OOMMF} M. J. Donahue and D. G. Porter, Interagency Report NO. NISTIR 6376, National Institute of Standards and Technology, Gaithersburg, MD (1999) [http://math.nist.gov/oommf/].

\bibitem{Koshibae_NCOMMS2016} W. Koshibae and N. Nagaosa, Theory of antiskyrmions in magnets, Nat. Commun. \textbf{7}, 10542 (2016).

\bibitem{Xichao_IEEE2017} X. Zhang, J. Xia, G.~P. Zhao, X. Liu, and Y. Zhou, Magnetic Skyrmion transport in a nanotrack with spatially varying damping and non-adiabatic torque, IEEE Tran. Magn. \textbf{53}, 1 (2017).

\bibitem{Thiele_PRL1973} A. A. Thiele, Steady-state motion of magnetic domains, Phys. Rev. Lett. \textbf{30}, 230 (1973).

\bibitem{Heyne_PRL2010} L. Heyne, J. Rhensius, D. Ilgaz, A. Bisig, U. R{\"u}diger, M. Kl{\"a}ui, L. Joly, F. Nolting, L. J. Heyderman, J. U. Thiele, and F. Kronast, Direct determination of large spin-torque nonadiabaticity in vortex core dynamics, Phys. Rev. Lett. \textbf{105}, 187203 (2010).

\bibitem{Pollard_NCOMMS2012} S. D. Pollard, L. Huang, K.S. Buchanan, D.A. Arena, and Y. Zhu, Direct dynamic imaging of non-adiabatic spin torque effects, Nat. Commun. \textbf{3}, 1028 (2012).

\bibitem{Reichhardt_NJP2016} C. Reichhardt and C. J. O. Reichhardt, Noise fluctuations and drive dependence of the skyrmion Hall effect in disordered systems, New J. Phys. \textbf{18}, 095005 (2016).

\bibitem{Senfu_NJP2015} S. Zhang, J. Wang, Q. Zheng, Q. Zhu, X. Liu, S. Chen, C. Jin, Q. Liu, C. Jia, and D. Xue, Current-induced magnetic skyrmions oscillator, New J. Phys. \textbf{17}, 023061 (2015).

\bibitem{LiuQF_PRAP2018} C. Jin, J. Wang, W. Wang, C. Song, J. Wang, H. Xia, and Q. Liu, Array of synchronized nano-oscillators based on repulsion between domain wall and skyrmion, Phys. Rev. Appl. \textbf{9}, 044007 (2018).

\end{thebibliography}
\end{document}